\documentclass[twocolumn,showpacs,preprintnumbers,amsmath,prb,amssymb,superscriptaddress,a4paper]{revtex4-1}

\pdfoutput=1

\usepackage[export]{adjustbox}
\usepackage{amsfonts}
\usepackage{amsmath}
\usepackage[makeroom]{cancel}
\usepackage[english]{babel}
\usepackage[T1]{fontenc}
\usepackage{times}
\usepackage{mathrsfs}
\usepackage{graphicx}
\usepackage{dcolumn}
\usepackage{bm}
\usepackage{wasysym}
\usepackage[colorlinks,bookmarks=true,citecolor=blue,linkcolor=red,urlcolor=blue]{hyperref}
\usepackage[tight, FIGTOPCAP, hang, raggedright, nooneline]{subfigure}
\usepackage{hyperref}
\usepackage{xcolor}
\usepackage{epsfig}
\usepackage{amssymb}
\usepackage{lipsum}
\usepackage{enumitem}

 \definecolor{Green}{RGB}{80,182,0}

\usepackage{color}

\newcommand{\be}{\begin{equation}}
\newcommand{\bea}{\begin{eqnarray}}
\newcommand{\ee}{\end{equation}}
\newcommand{\eea}{\end{eqnarray}}

\newcommand{\Sbf}{\mathbf{S}}
\newcommand{\rbf}{\mathbf{r}}
\newcommand{\gradbf}{\mbox{\boldmath $\nabla$}}
\newcommand{\dotbf}{\mbox{\boldmath $.$}}

\begin{document}

\title{Zero point fluctuations for magnetic spirals and Skyrmions,\\ and the fate of the Casimir energy in the continuum limit}

\author{B. Dou\c{c}ot}
\affiliation{LPTHE, CNRS and Universit\'e Pierre and Marie Curie, Sorbonne Universit\'es, 75252 Paris Cedex 05, France}
\author{D.~Kovrizhin}
\affiliation{Rudolf Peierls Centre for Theoretical Physics, Oxford University, Oxford OX1 3PU, United Kingdom}
\affiliation{NRC Kurchatov institute, 1 Kurchatov square, 123182, Moscow, Russia}
\author{R. Moessner}
\affiliation{Max-Planck-Institut f\"ur Physik komplexer Systeme, 01187 Dresden, Germany}

\date{\today}

\begin{abstract}
We study the role of zero-point quantum fluctuations in a range of magnetic states which on the classical level are close to  spin-aligned ferromagnets. These include Skyrmion textures characterized by non-zero topological charge, and topologically-trivial spirals arising from the competition of the Heisenberg and Dzyaloshinskii-Moriya interactions. For the former, we extend our previous results on quantum exactness of classical Bogomolny-Prasad-Sommerfield (BPS) ground-state degeneracies  to the general case of K\"ahler manifolds, with a specific example of Grassmann manifolds $\mathrm{Gr(M,N)}$. These are relevant to quantum Hall ferromagnets with $\mathrm{N}$ internal states and integer filling factor $\mathrm{M}$. A promising candidate for their experimental implementation is monolayer graphene with $\mathrm{N=4}$ corresponding to spin and valley degrees of freedom at the charge neutrality point with $\mathrm{M=2}$ filled Landau levels. We find that the vanishing of the zero-point fluctuations in taking the continuum limit occurs differently in the case of BPS states compared to the case of more general smooth textures, with the latter exhibiting more pronounced lattice effects. This motivates us to consider the vanishing of zero-point fluctuations in such near-ferromagnets more generally.  We present a family of lattice spin models for which the vanishing of zero-point fluctuations is evident, and show that some spirals can be thought of as having nonzero but weak zero-point fluctuations on account of their closeness to this family. Between them, these instances provide concrete illustrations of how the Casimir energy, dependent on the full UV-structure of the theory, evolves as the continuum limit is taken. 
\end{abstract}

\maketitle

\section{Introduction}
The classical ground state of a Heisenberg ferromagnet, which has all its spins
co-aligned, is well-known to be special in that it is not at all dressed by quantum
fluctuations. In other words this ground state is an eigenstate of the quantum
Hamiltonian for any spin-length. Recently we showed \cite{DKM16} that this very special
property also holds in certain {\it non-collinear} spin-textures in quantum ferromagnets.
The latter, which arise as analytic topological solutions in the framework of the analysis
of a non-linear sigma model, remain undressed to all orders of the semiclassical expansion
based on a \textit{geometric quantization} scheme \cite{Perelomov}. This offers
a new perspective on the physics of two-dimensional non-collinear quantum magnets,
in particular in relation to exactness of quantum degeneracies of BPS
states found in the studies of supersymmetric sigma-models \cite{Olive}.

Our original observation \cite{DKM16} immediately raised a number of
interesting questions. First, what is the manifestation of this phenomenon in the more
familiar (and pedestrian) setting of the semiclassical spin-wave
treatment of such magnets. Second, given that our previous calculations were undertaken directly in the continuum limit, how does the presence of finite lattice effects modify these results. Third, and most fundamentally, 
does our observation represent a singular, and therefore rather special, instance of the behaviour of
a quantum magnet, or perhaps one can find an underlying structure responsible for this phenomenon
that can be used to extend our findings to a generalised class of models. 

In this work, we address all of these items. As we rely on a number of rather different theoretical
approaches and methods, following this introduction we subdivide the remainder of this account into
three, more or less, independent sections which are intended to be largely self-contained. The first two
are devoted to complementary accounts of Skyrmion magnets of the type encountered in quantum Hall physics.
The third, in turn, provides a juxtaposition of these ideas to the  setting of `conventional' spiral phases
induced by magnetic frustration or spin-orbit coupling, demonstrating that the absence of zero-point
fluctuations is perhaps more widespread than might have been naively expected.

In the first part of the paper we apply Holstein-Primakoff spin-wave theory to the lattice version
of topologically-nontrivial ferromagnetic spin textures whose continuum limit was found previously
to be exact quantum mechanically. The lattice Hamiltonian we consider is that of a simple Heisenberg ferromagnet. 
We study the spin-wave expansion for the latter on a sphere and on a torus, where these topologically nontrivial textures  can be compactly parametrised. First of all, we find that such textures, even though they 
are not the ground states of the quantum lattice Hamiltonian (the uniform ferromagnetic state is),
are locally stable, as the spin rearrangement which is necessary to reach the ground state is topological and
therefore cannot be accessed via local deformations captured by  spin wave theory. Further we show that,
to leading order in the spin-wave expansion, the anomalous terms responsible for quantum dressing
such as appearing in the case of a Heisenberg antiferromagnet, vanish in the continuum limit.  

We supplement the spin-wave analytics  with a detailed numerical implementation of spin-wave theory
on a sphere and a torus. For the case of the sphere, we introduce a lattice  via a triangulation procedure, 
and extract the spectrum of the Hessian around such a Skyrmion texture on the resulting  triangulated spheroidal surface. This turns out to take the form of the spectrum of Landau levels on the sphere,
i.e.~the texture acts as if to supply the charge of a monopole sitting at the centre of the sphere.
The spectrum gains some additional fine structure, compared to the idealised spectrum in the continuum case,
on account of the fact that the triangulations (necessarily though weakly) break the full rotational symmetry of the sphere.
By  varying the density of the triangulation, we can study finite-size effects beyond the continuum limit. 
We find that the zero-point energy vanishes inversely proportional to the square of the number of lattice sites
if the continuum limit of the texture belongs to the BPS manifold, and in this case the vanishing occurs parametrically faster than for the case of a general smooth texture having components in the non-analytic subspace. 

In the second part of this paper we adopt a perspective amenable to abstraction and generalisation
based on the standard treatment of an ordered ferromagnet within a framework of nonlinear sigma-model. In our 
earlier work \cite{DKM16} we noted that it was the analytical nature of the minimal-energy spin-textures
which led to the absence of the zero-point fluctuations and dressing. This goes along with  the saturation
of the BPS bound, which states that the energy of a topological texture is at least as large as its
topological charge (with a known proportionality constant). The BPS bound is saturated if and only if the texture is (anti-)holomorphic \cite{Rajaraman}. 

The central step in the generalisation of this result, that we present in this paper, lies in the identification
of this structure as a specific instance of the defining property of a K\"ahler manifold.
A K\"ahler manifold is a complex manifold equipped with a metric which can locally
be obtained by differentiating a K\"ahler potential \cite{Arnold}. The existence of the potential can now  
be used to generalise the Bogomolny inequality to all textures which map  real space into a the K\"ahler manifold
in an (anti-)analytic manner for positive (negative) topological charge. Our previous arguments then carry over more or less directly to this generalised setting. This widens their remit to include other such manifolds, e.g.~in the hitherto unexplored case of topological spin textures of quantum Hall ferromagnets away from unit filling; this includes the case of graphene in a magnetic field near the charge neutrality point. 

We then turn to the question whether this phenomenon of vanishing zero-point fluctuations is a general feature
of slowly twisted `almost-ferromagnets' or it is a very special feature of spin-textures living in the BPS manifold.
To this effect we analyze a well-known example of a quantum one-dimensional magnet whose classical ground state
is a spin-spiral induced by a spin-orbit coupling, such as in the example of Dzyaloshinskii-Moriya (DM) interactions (we find that competing further-neighbour Heisenberg interactions produce a similar result). 
We find that this indeed also has non-vanishing zero-point fluctuations in the continuum limit. However, in the (often physically relevant) discrete lattice incarnation, zero-point fluctuations do not vanish.  

We show that there do in fact exists a family of lattice Hamiltonians for which the zero-point fluctuations vanish. These are obtained from a 'standard' ferromagnet via a unitary transformation preserving the full spectrum, and hence the absence of dressing. A prominent recent example of this is provided by a special point of the Heisenberg-Kitaev model \cite{Jackeli-Khalliulin}. We find that the DM spiral is close to this model, in the sense that it approaches such a special point in the continuum limit, hence accounting naturally for the nonzero but small zero-point fluctuations en route. 

We close the manuscript with a summary and some thoughts on broader implications of this work.

\section{Quantum spin-textures with topological charge}

\subsection{General considerations}

Consider a Heisenberg Hamiltonian with a ferromagnetic exchange, $J>0$, with spins defined on a locally-planar graph
\begin{equation}\label{spin_hamiltonian}
\hat{H}=-J\sum_{\left\langle ij \right\rangle}\hat{\mathbf{S}}_{i}\hat{\mathbf{S}}_{j}.
\end{equation}
At each lattice site $i$, we attach a spin-operator $\hat{\mathbf{S}}_i$ belonging to the spin-$S$ representation of the $\mathrm{SU(2)}$.
In the large $S$ limit it
is also possible to define spin-coherent configurations associated to classical textures carrying any chosen integer topological
charge. Such a classical texture is defined by two angular fields $\theta_i\in [0,\pi]$ and $\phi_i \in [0,2\pi)$, such that the spin coherent state on site $i$ can be parametrised by a three-component vector of unit length
\begin{equation}
\mathbf{S}_{i}=(\sin \theta_{i}\cos \phi_{i}, \sin \theta_{i}\sin \phi_{i},\cos \theta_{i}).  
\end{equation}
The corresponding semiclassical wave-function is given by a product state
\begin{equation}
\left |\Psi \right\rangle=\otimes_{i}\left |\mathbf{S}_{i}\right\rangle,
\end{equation} 
where $\left |\mathbf{S}_{i}\right\rangle$ denotes the spin coherent state on site $i$. The latter can be viewed
as a $2S$-fold symmetrized tensor product of the standard spin-$1/2$ spinor 
$\left |\theta_{i}\ \phi_{i}\right\rangle$ defined by
\begin{equation}
\left |\theta_{i}\ \phi_{i}\right\rangle =
\cos(\theta_{i}/2) \left |\uparrow \right\rangle_{i}+e^{i\phi_{i}}\sin(\theta_{i}/2)\left |\downarrow \right\rangle_{i}.
\end{equation}
Now, we perform a unitary transformation on the spin Hamiltonian (\ref{spin_hamiltonian}) by locally rotating it into a frame co-aligned with the classical spin orientation on every site. 

Denoting by $\hat{\mathbf{T}}_{i}$ the operator of the spin at site $i$ in this rotated frame, we set
$\hat{\mathbf{S}}_{i}=\mathcal{R}_{\mathbf{z}}(\phi_{i})\mathcal{R}_{\mathbf{y}}(\theta_{i})\hat{\mathbf{T}}_{i}$, which reads:
\begin{eqnarray*}
\hat{S}_{i}^{z} &=& \hat{T}_{i}^{z}\cos\theta_{i}-\frac{1}{2}(\hat{T}_{i}^{+}+\hat{T}_{i}^{-})\sin \theta_{i}, \\
\hat{S}_{i}^{+} &=&  [\hat{T}_{i}^{z}\sin \theta_{i}+\cos^{2}(\theta_{i}/2)\hat{T}_{i}^{+}-\sin^{2}(\theta_{i}/2)\hat{T}_{i}^{-}]e^{i\phi_{i}}, \\
\hat{S}_{i}^{-} &=&  [\hat{T}_{i}^{z}\sin \theta_{i}-\sin^{2}(\theta_{i}/2)\hat{T}_{i}^{+}+\cos^{2}(\theta_{i}/2)\hat{T}_{i}^{-}]e^{-i\phi_{i}}.             
\end{eqnarray*}
In the rotated frame, the classical texture becomes a ferromagnetic state aligned along the vertical axis in spin space.
As usual, it is convenient to quantize the spin system using Holstein-Primakoff (HP) bosons $\hat{b}_{i}$,  $\hat{b}^{\dagger}_{i}$,
\begin{equation*}
\hat{T}_{i}^{z} =  S- \hat{b}^{\dagger}_{i}\hat{b}_{i},\ 
\hat{T}_{i}^{+} = [2S- \hat{b}^{\dagger}_{i}\hat{b}_{i}]^{\frac{1}{2}}\hat{b}_{i},\ 
\hat{T}_{i}^{-} =  \hat{b}^{\dagger}_{i}[2S- \hat{b}^{\dagger}_{i}\hat{b}_{i}]^{\frac{1}{2}}.    
\end{equation*}
In the large $S$ limit, and to leading order in $1/S$ expansion, it is sufficient for our purposes to keep the following terms:
\begin{equation*}
\hat{T}_{i}^{z} = S-\hat{b}^{\dagger}_{i}\hat{b}_{i},\ 
\hat{T}_{i}^{+} = \sqrt{2S}\hat{b}_{i},\ 
\hat{T}_{i}^{-} = \hat{b}^{\dagger}_{i}\sqrt{2S}.    
\end{equation*}
Substituting these expressions into the Heisenberg Hamiltonian leads to the expansion:
$\hat{H}=E_{0}+\hat{H}_{1}+\hat{H}_{2}+...$ The first term $E_{0}$ is the classical energy of the spin-texture. Defining 
\begin{equation*}
s_i\equiv \sin\theta_i,\ c_i\equiv \cos\theta_i,\ \phi_{ij}\equiv \phi_i-\phi_j,\ \theta_{ij}\equiv \theta_i-\theta_j
\end{equation*}
we can write
\begin{equation} 
E_{0}=-JS^{2}\sum_{\left\langle ij \right\rangle}(c_i c_j+s_i s_j\cos\phi_{ij}).
\end{equation}
The next term $\hat{H}_{1}$ is linear in HP boson operators
\begin{equation*}
\hat{H}_{1}=\frac{JS^{\frac{3}{2}}}{\sqrt{2}}\sum_{i,j\in \mathcal{N}_i}\hat{b}_{i}(s_{i}c_{j}
-c_{i}s_{j}\cos\phi_{ij}-i s_{j}\sin\phi_{ij})+h.c.
\end{equation*} 
Here $\mathcal{N}_i$ denotes the set of nearest neighbours of site $i$.
Note that this linear term vanishes in extremal configurations for the classical energy, i.e.~those configurations for which
both $\partial E_{0}/\partial \theta_{i}=0$ and  $\partial E_{0}/\partial \phi_{i}=0$.
In such situation, the next correction, $\hat{H}_{2}$, which is quadratic in HP boson operators, is particularly important
because it determines the magnon spectrum in the vicinity of the classical coherent state, and also the first
quantum corrections to its zero point energy. We have
\begin{align*}
\hat{H}_{2} {}&= \frac{JS}{2}\sum_{\left\langle ij \right\rangle}(c_{i}c_{j}+s_{i}s_{j}\cos\phi_{ij})
(\hat{b}^{\dagger}_{i}\hat{b}_{i}+\hat{b}^{\dagger}_{j}\hat{b}_{j}) \\
-{}& [s_{i}s_{j}+
(1+c_{i}c_{j})\cos\phi_{ij}-i(c_{i}+c_{j})\sin\phi_{ij}]\hat{b}^{\dagger}_{i}\hat{b}_{j} \\
-{}&[s_{i}s_{j}+
(c_{i}c_{j}-1)\cos\phi_{ij}-i(c_{i}-c_{j})\sin\phi_{ij}]\hat{b}_{i}\hat{b}_{j}+h.c. 
\end{align*}
One can write this quadratic Hamiltonian $\hat{H}_2$ in the standard form \cite{Blaizot}
\begin{equation}
 \hat{H}_{2}= \sum_{\left\langle ij \right\rangle} A_{ij}\hat{b}^{\dagger}_{i}\hat{b}_{j}+\frac{1}{2}(B_{ij}\hat{b}^{\dagger}_{i}\hat{b}^{\dagger}_{j}+B_{ij}^{\ast}\hat{b}_{i}\hat{b}_{j}),
\end{equation}
where the $N \times N$ matrix $A$ is hermitian and $B$ is symmetric. After Bogoliubov transformation, see \cite{Blaizot}, the quadratic Hamiltonian can be written in a diagonal form
\begin{equation}
\hat{H}_{2}= \sum_{n=1}^{N}\omega_{n}\hat{\beta}_{n}^{\dagger}\hat{\beta_{n}}+\delta,
\end{equation}
where $\omega_n$ are the eigenfrequencies of the Hamiltonian, and
\begin{equation}
 \delta= \frac{1}{2}\sum_{n=1}^{N}\omega_{n}-\frac{1}{2}\mathrm{Tr}A
\end{equation}
is the correction to the ground-state energy due to quantum fluctuations, up to first non-trivial order in the $1/S$ expansion.
In the quantum magnetism literature, the ground-state energy $E_{\mathrm{GS}}$ is often written as
\begin{equation}
E_{\mathrm{GS}}= \tilde{H}_{0}+\frac{1}{2}\sum_{n=1}^{N}\omega_{n},
\end{equation}
where $\tilde{H}_{0}$ is obtained after replacing $S^{2}$ by the quantum mechanically dressed $\mathbf{S}^{2}=S(S+1)$ in the
prefactor of the classical energy $E_{0}$. We prefer not to use this traditional form here, because the correction term $\delta$
vanishes  whenever the elements of the matrix $B_{ij}=0$, which is of course the case for uniform classical configurations.
The presence of spatial gradients in a {\em generic} texture generates a non-zero $B$ matrix, which in turn modifies the ground-state energy compared to its classical value $E_{0}$.

It should be noted that $A_{ii}$, $A_{ij}$, and  $B_{ij}$ are not independent. If we write $A_{i}\equiv\sum_{j\in \mathcal{N}_i}A_{ii}^{(j)}$, then
\begin{equation} 
|A_{ij}|+|B_{ij}|  =  JS,\ |A_{ij}|-|B_{ij}|  =  |A_{i}^{(j)}|.
\end{equation}
For smooth textures, we can simplify the expressions of the $A$ and $B$ matrices, assuming that $\theta_{ij}$
and $\phi_{ij}$ are small, being of the order of the lattice spacing $a$. Defining $\Theta_{ij}=(\theta_i+\theta_j)/2$ we have for the diagonal terms of the matrix $A$
\begin{equation}
A_{ii}\simeq JS \sum_{j \in \mathcal{N}_i}(1-\frac{1}{2}\theta_{ij}^{2}-\frac{1}{2}\phi_{ij}^{2}\sin^{2}\Theta_{ij}),
\end{equation}
and similarly for the off-diagonal terms of $A$ and $B$
\begin{eqnarray}
A_{ij}  & \simeq&  -JS (1-\frac{1}{4}\theta_{ij}^{2}-\frac{1}{4}\phi_{ij}^{2}\sin^{2}\Theta_{ij})
e^{-i\phi_{ij}\cos\Theta_{ij}},\nonumber
\\
B_{ij}  &\simeq&  \frac{JS}{4}
(\theta_{ij}+i\phi_{ij}\sin\Omega_{ij})^{2}.
\label{approx_Bij}
\end{eqnarray}
One can see that for smooth textures, the entries of the matrix $A_{ij}$ (related to the coefficients of $\hat{b}^{\dagger}_{i}\hat{b}_{j}$ terms) remain finite in the continuum limit when the lattice spacing $a$ sent to zero, but the entries of the matrix $B$ (related to the coefficients of the Bogoliubov terms $\hat{b}_{i}\hat{b}_{j}$) are quadratic 
in the deviations between angles at neighbouring sites.

In the rest of the paper we shall focus on the behaviour of the quantum corrections to ground-state energy in the
continuum limit $a \rightarrow 0$, so we need to specify more precisely how we take this limit. The guiding principle is that
we scale the magnetic coupling $J$ in such a way that the classical term $E_0$ becomes a smooth local functional of the unit vector field $\Sbf(\rbf)$ which interpolates between the values $\Sbf_{i}$ on the lattice sites $i$.

For a regular square lattice in a $D$-dimensional space, we need then to take $J$ proportional to $a^{D-2}$. This ensures that in the continuum limit the difference $E_0-E_{f}$, where $E_{f}=-J S^2 N_b$ is the energy of a ferromagnetic state for a system with $N_b$ bonds, is proportional to $\int d^{D}x\ (\gradbf \Sbf)^{2}$. In other words we scale the lattice spacing and the coupling in such a way that the classical energy of the topological spin-texture attains a constant value in the continuum limit. In most of the paper, we shall work in $D=2$ dimensions, and therefore $J$ takes a fixed value independent of $a$.

\begin{figure}[b]
\centering
\includegraphics[width=5cm]{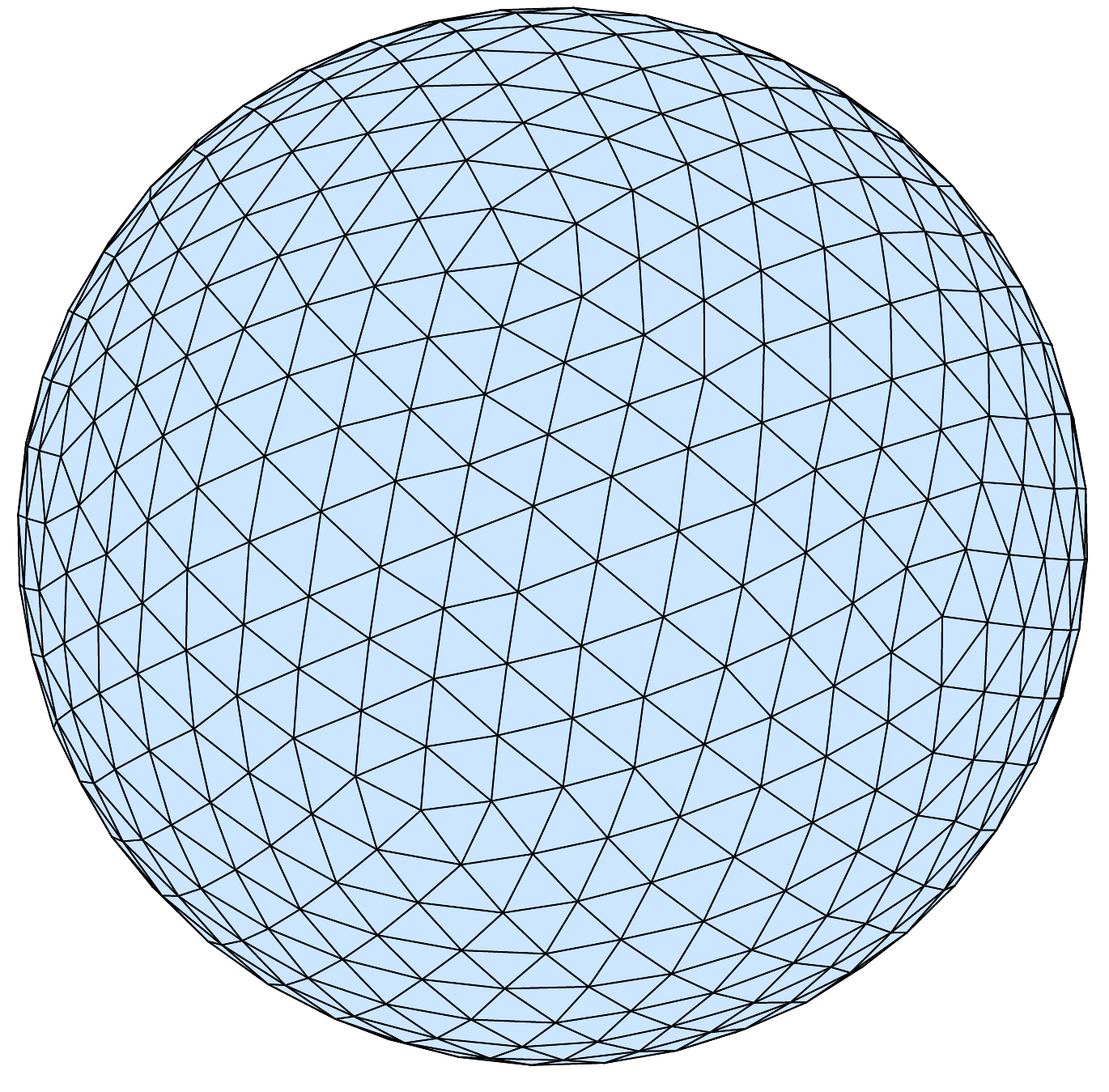}
\caption{An example of a triangulation of a sphere (here shown with 642 lattice sites). In the studies of the continuum limit of a Heisenberg model on a sphere we use triangulations with increasing number of lattice sites. The spins are arranged on the sites. Note the presence of defects, e.g.~pentagonal loops.}\label{fig:triang}
\end{figure}

\subsection{Single-magnon hopping}

It is first interesting to study the single magnon hopping term. Neglecting second order corrections
in spin gradients, we can approximate $A_{ij}\simeq-JS\exp(-i\mathcal{A}_{ij})$, where 
$\mathcal{A}_{ij}=\phi_{ij}\cos\Omega_{ij}$. We see that in the continuum limit magnons experience
an effective static orbital magnetic field described by the vector potential
$\mathbf{\mathcal{A}}=\cos(\theta)\gradbf \phi$. This is a manifestation of the Berry phase accumulated by a
magnon moving along a closed path $\gamma$ in a slowly varying spin-background. This Berry phase is equal to
the solid angle $\Omega_{\gamma}$ spanned by the spin texture on the unit sphere along $\gamma$, multiplied by the
spin of the magnon. A direct calculation shows that $\oint_{\gamma}\mathbf{\mathcal{A}}\mathbf{.}d\mathbf{l}=\Omega_{\gamma}$,
which is consistent with the fact that a magnon is a spin-1 object. Because  $\Omega_{\gamma}$ is equal to $4\pi$ times the
number of Skyrmions enclosed in $\gamma$, the effective magnetic flux seen by the magnon is twice the number of Skyrmions
inside $\gamma$.

\begin{figure}[t]
\centering
\includegraphics[width=\columnwidth]{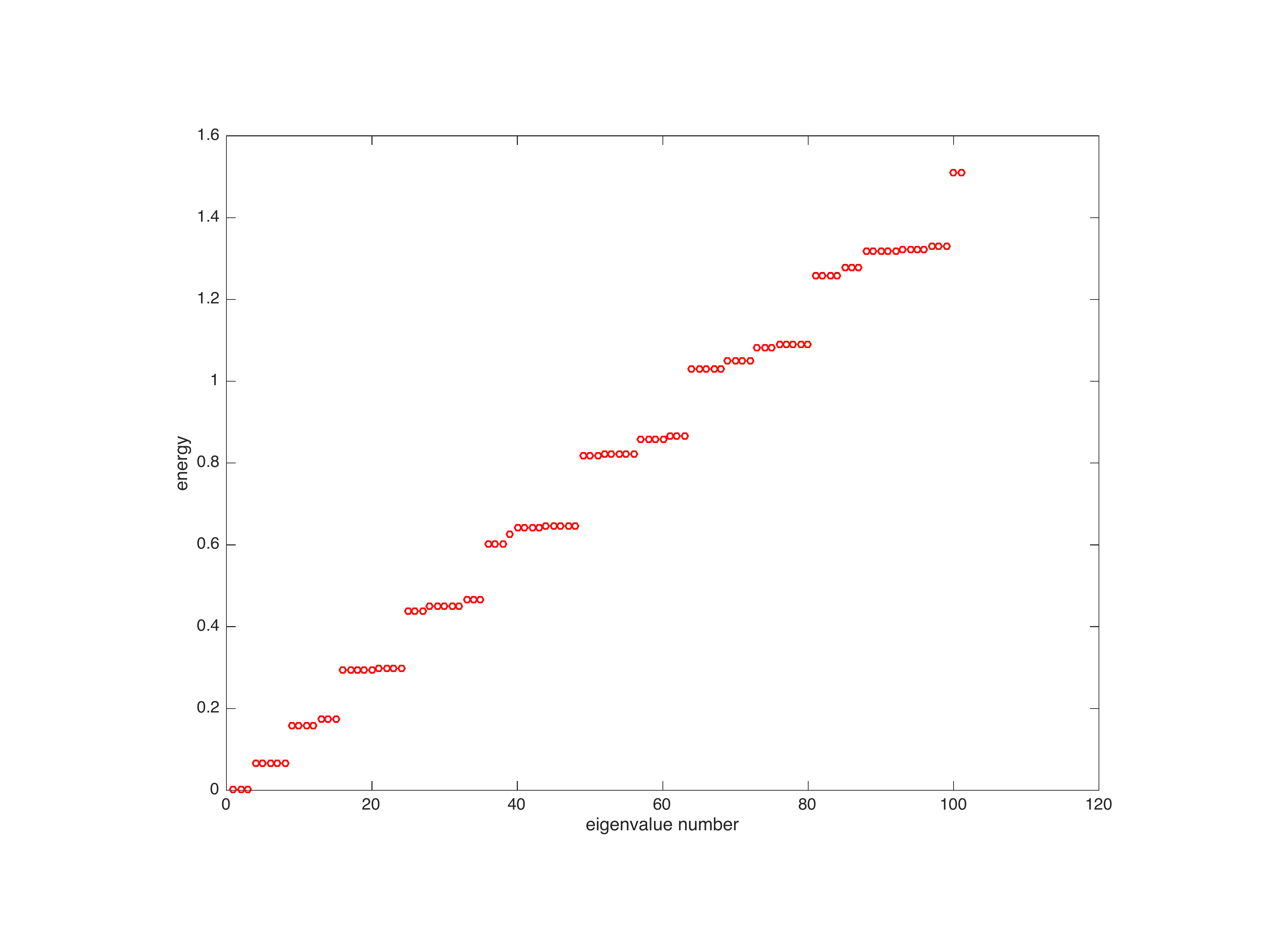}
\caption{Landau levels from the diagonalization of the Hamiltonian $\hat{H}_2$ in the presence of a Skyrmion state on a sphere. The degeneracies are in agreement with the results obtained for the spectrum of a Schr\"odinger equation in the presence of a monopole, see~\cite{Haldane83}.}\label{fig:LL}
\end{figure}

Our numerical results for various triangulations on a sphere, in the background of a single Skyrmion,
are fully consistent with these considerations. Indeed, for a single particle moving on a sphere enclosing $2s$ magnetic
monopoles ($s$ is integer or half-integer) we have a Landau level spectrum with eigenenergies $E_n$ ($n \geq 0$) 
proportional to $n^{2}+(2s+1)n+s$, the $n$-th level being $2n+2s+1$-fold degenerate~\cite{Haldane83}.

Fig.~2 shows the low-lying spectrum of the quadratic $\hat{H}_2$ correction, (where magnon number non-conserving terms have also been included), for a symmetric triangulation of the sphere with 642 sites, see Fig.~1. (The triangulations are constructed using two different algorithms. The first algorithm performs successive subdivision of a sphere into triangles (think e.g.~of a surface of a golf-ball), while the second algorithm is based on energy minimization for a system of points connected by springs for a given number of points.) The Landau levels compatible with $s=1$ (i.e. two monopoles inside the sphere) are clearly visible in Fig.~2, from $n=0$ to $n=8$. Degeneracies are lifted by residual rotational symmetry breaking terms induced by the triangulation, but the effect of such terms appears to be rather small. The three degenerate zero modes arise from the spontaneous breaking of spin rotation symmetry in a non-coplanar coherent state.

Further support to the picture of magnons evolving in a smooth magnetic field is given by the scaling of the Landau level gaps, which are proportional to $N^{-1}$, as show on Fig.~\ref{fig:scaling_sphere}. In the large $N$ limit, the $A$ operator is proportional to $a^{2}$ times a discretized Laplacian, in the presence of a fixed magnetic field. The spectrum of this discretized Laplacian has gaps which reach a finite limiting value in the large $N$ limit. Since $Na^{2}$ is constant, we expect Landau level gaps for the magnon excitations to be proportional to $N^{-1}$ at large $N$.

\subsection{Effect of magnon non-conserving terms}

Our general discussion, together with the excellent agreement between the numerical magnon spectrum and
predictions based only on the magnon number conserving term ($A$ operator), suggest that it is legitimate to
treat the $B$ terms as a small perturbation. Eigenmodes $\hat{\beta}^{\dagger}$ of $\hat{H}_2$ are obtained using equations of motion $[\hat{H}_{2},\hat{\beta}^{\dagger}]=\omega\hat{\beta}^{\dagger}$. Writing $\hat{\beta}^{\dagger}=\sum_{i=1}^{N}(u_{i}\hat{b}^{\dagger}_{i}+v_{i}\hat{b}_{i})$, we see
that the $2N$-component vector $|\psi\rangle = (u_1,...u_N,v_1,...,v_N)^{T}$ is an eigenvector of the $2N$ times $2N$
matrix M defined as
\begin{equation}
M=\left(\begin{array}{lr}
A & -B \\
\bar{B} & -\bar{A} 
\end{array}\right)
\end{equation}
This matrix $M$ is self-adjoint for the indefinite hermitian form in $\mathbb{C}^{2N}$ associated to the matrix K, 
so $M^{\dagger}K=KM$, with
\begin{equation}
K=\left(\begin{array}{lc}
I_{N} & 0 \\
0 & - I_{N} 
\end{array}\right)
\end{equation}
This can be  checked from the definitions of $M$ and $K$, and is directly related to the fact that canonical
commutators are preserved under the time evolution generated by the Hamiltonian $\hat{H}_2$, see \cite{Blaizot}. 

\begin{figure}[b]
\centering
\includegraphics[width=\columnwidth]{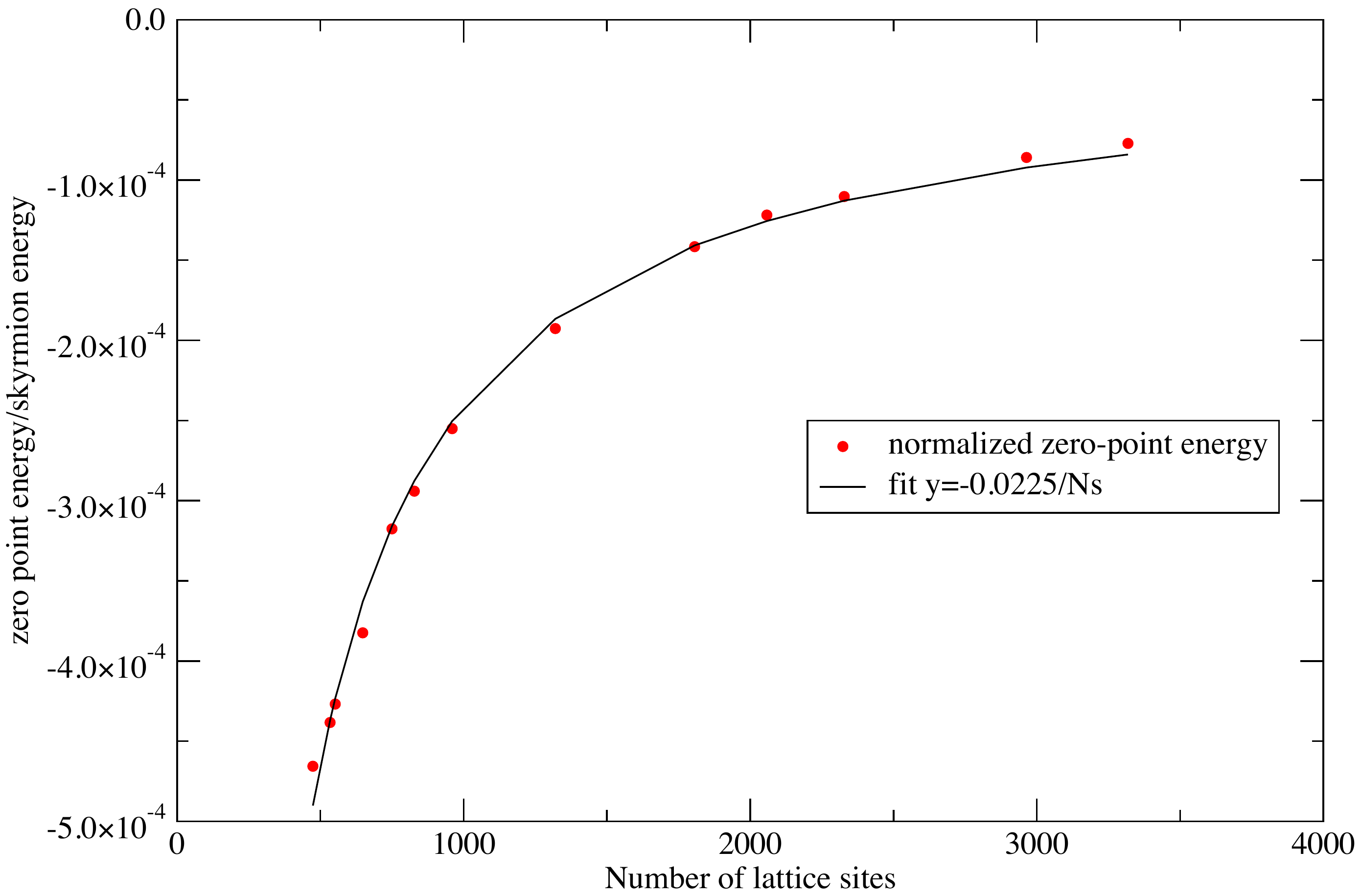}
\caption{Scaling of zero-point energy for a Skyrmion state on a sphere as a function of number of lattice sites $N_s$. (red dots) zero point energy obtained from numerical diagonalization of the Hamiltonian $\hat{H}_2$ in the background of a classical Skyrmion spin-texture. (solid line) fit with $const/N_s$.}\label{fig:scaling_sphere}
\end{figure}

It should be noted that $[\hat{H}_{2},\hat{\beta}^{\dagger}]=\omega\hat{\beta}^{\dagger}$ implies that
$[\hat{H}_{2},\hat{\beta}]=-\omega\hat{\beta}$, so all non-zero eigenvalues of $M$ can be grouped into opposite pairs. Creation operators $\beta^{\dagger}_{n}$ are then associated to positive eigenvalues $\omega_{n}$. See a careful discussion of possible zero-modes in the general case \cite{Blaizot}, and also the example of DM-spiral below.

Let us decompose $M$ into $M_0+M_1$, where $M_0$ involves only $A$ and $M_1$ only $B$. We consider the $n$-th eigenvector
$|\psi_{n}\rangle$ of $M$ with positive eigenvalue $\omega_{n}$. When $B=0$, the $v$ components of $|\psi_{n}^{(0)}\rangle$  
are all equal to zero. Applying $M_1$ to $|\psi_{n}^{(0)}\rangle$ produces a new vector which is $K$-orthogonal to  $|\psi_{n}^{(0)}\rangle$,
so the first order correction to  $\omega_{n}^{(0)}$ vanishes. Applying standard perturbation theory (the only special feature here is the indefinite nature of the hermitian product $K$) gives the second order correction:
\begin{equation}
\delta \omega_{n} = -\frac{\langle u_{n}^{(0)}|B(\omega_{n}^{(0)}+\bar{A})^{-1}\bar{B}| u_{n}^{(0)}\rangle}
{\langle u_{n}^{(0)}|u_{n}^{(0)}\rangle}
\label{2_order_correction}
\end{equation}
Here $|u_{n}^{(0)}\rangle$ denotes the $N$ component vector built from the $u$ components of $|\psi_{n}^{(0)}\rangle$. Because the 
matrix $A$ is positive definite, $\delta \omega_{n}<0$, and we have
\begin{equation}
|\delta \omega_{n}| < \frac{\langle u_{n}^{(0)}|B\bar{B}| u_{n}^{(0)}\rangle}{\langle u_{n}^{(0)}|u_{n}^{(0)}\rangle}
\frac{1}{\omega_{n}^{(0)}} \leq \frac{c a^{4}}{\omega_{n}^{(0)}}
\end{equation}
where $c$ is a constant of order unity. To second order in $B$ the quantum correction to the ground-state energy is
$\delta=\sum_{n=1}^{N}\delta \omega_{n}$. It is negative, and $|\delta|< ca^{4}\sum_{n=1}^{N}\frac{1}{\omega_{n}^{(0)}}$.
Note that the zero modes are not expected to appear in the second order correction Eq.~(\ref{2_order_correction}),
because these zero modes form an isolated subspace which is decoupled from the positive energy modes,  as a consequence of spontaneously broken rotational symmetry in the presence of a Skyrmion. However, these zero-modes do produce a non-zero contribution to the zero-point energy on the lattice in the case of DM-spiral state.
\begin{figure}[t]
\centering
\includegraphics[width=0.7\columnwidth]{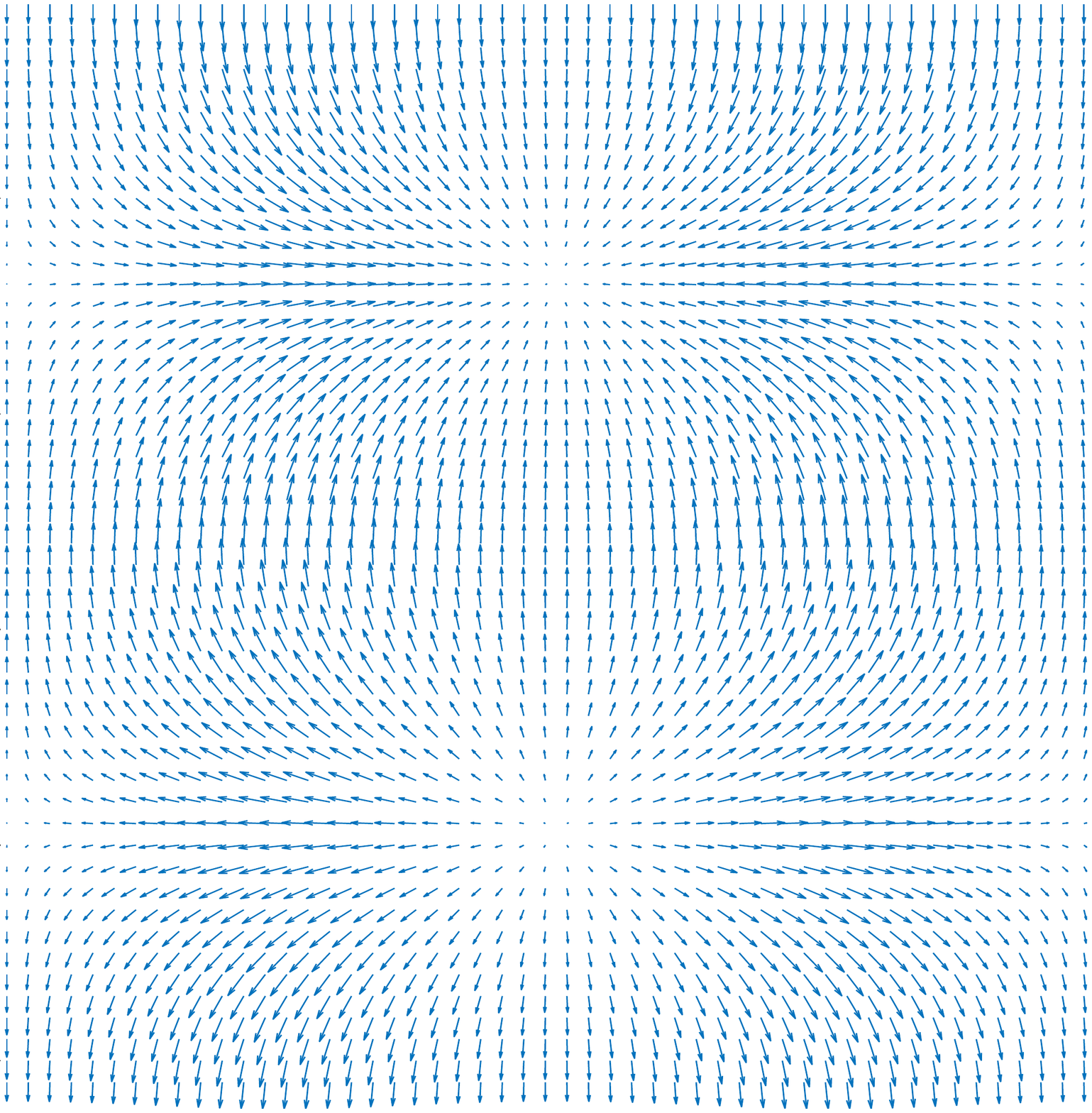}
\caption{Spin-texture for a Skyrmion on a torus}\label{fig:skyrmion_torus}
\end{figure}

\begin{figure}[t]
\centering
\includegraphics[width=0.95\columnwidth]{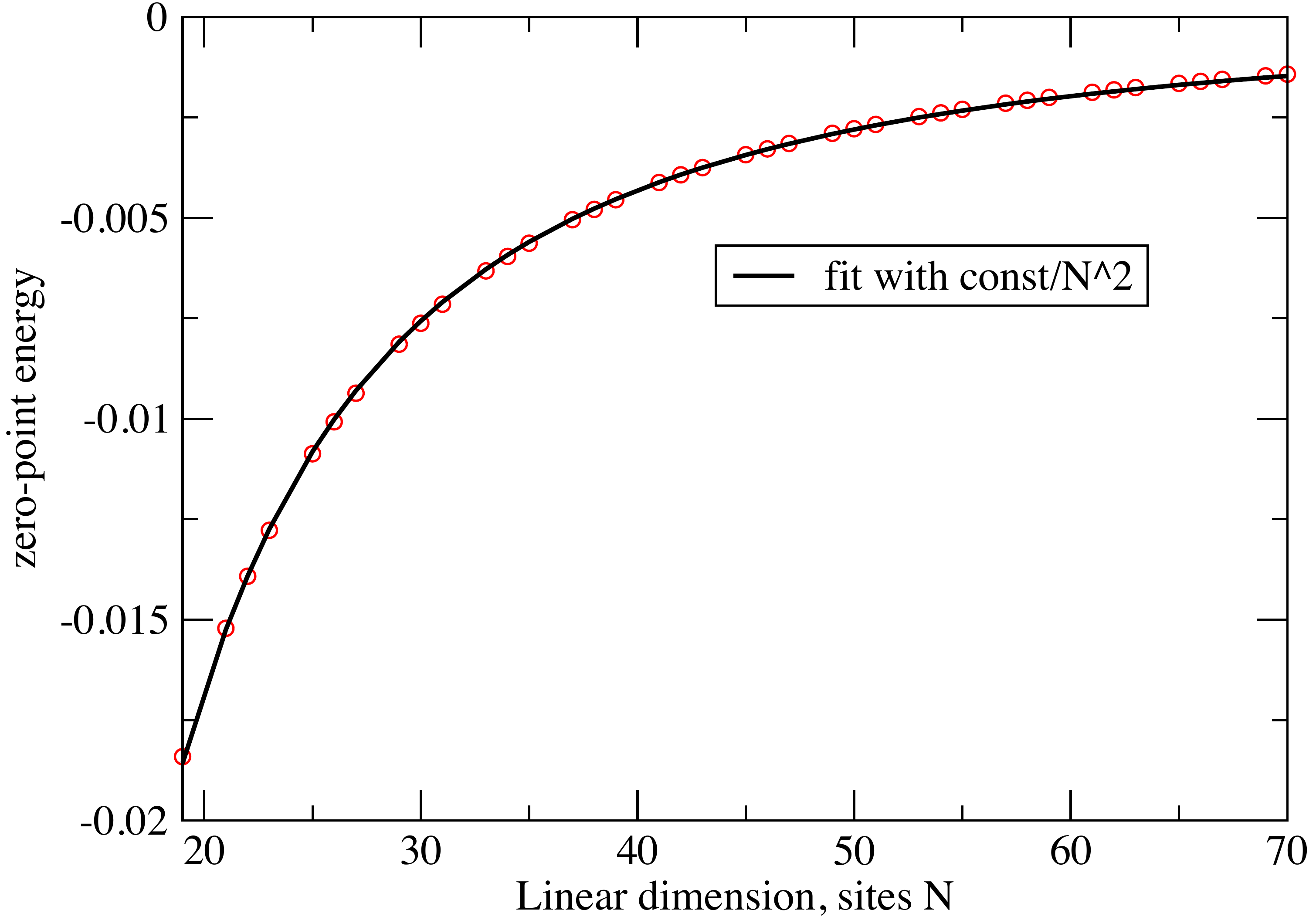}
\caption{Scaling of zero-point energy for a Skyrmion state on a torus as a function of linear number of sites $N$. (red dots) zero point energy obtained from numerical diagonalization of the Hamiltonian $\hat{H}_2$ in the background of a classical Skyrmion spin-texture. (solid line) fit with $const/N^2$.}\label{fig:scaling_torus}
\end{figure}

Since the average density of states for a $2D$ Laplacian is approximately constant, and since the 
Landau gap scales as $1/N$, we expect the sum over inverse eigenvalues of $A$ to scale like $N\log N$. Therefore, 
$|\delta|<c'N^{-1} \log N$, and $\delta \rightarrow 0$ as $N \rightarrow \infty$: the quantum correction 
to the ground-state energy due to $\hat{H}_2$ vanishes in the large $N$ limit for any smooth texture that is a local
minimum of the Heisenberg ferromagnetic Hamiltonian. This expectation is confirmed by the numerical results shown on Figs.~\ref{fig:scaling_sphere},\ref{fig:scaling_torus},
which suggest that $\delta$ is simply proportional to $1/N$. We have not examined the possible presence of 
logarithmic corrections, whose existence might be suggested by the upper bound just discussed.

So far, we have not made any use of the fact that, in the continuum limit, optimal Skyrmion textures are analytic 
functions of the complex position variable $z$ (which is obtained by stereographic projection of the sphere
onto the complex plane). By contrast, analyticity considerations played a key role in our previous work~\cite{DKM16}
and its generalization to arbitrary K\"ahler order parameter manifolds to be discussed in section~\ref{Generalized_to_Kaehler} below. There is no contradiction here, because analyticity has not been claimed to be a {\it necessary} condition for the suppression of quantum corrections to Skyrmion energy. However, analyticity has  farther-reaching consequences, and in particular, it preserves the classical texture state from any quantum corrections, to {\it all orders} in a $1/S$ expansion.

\begin{figure}[b]
\centering
\includegraphics[width=\columnwidth]{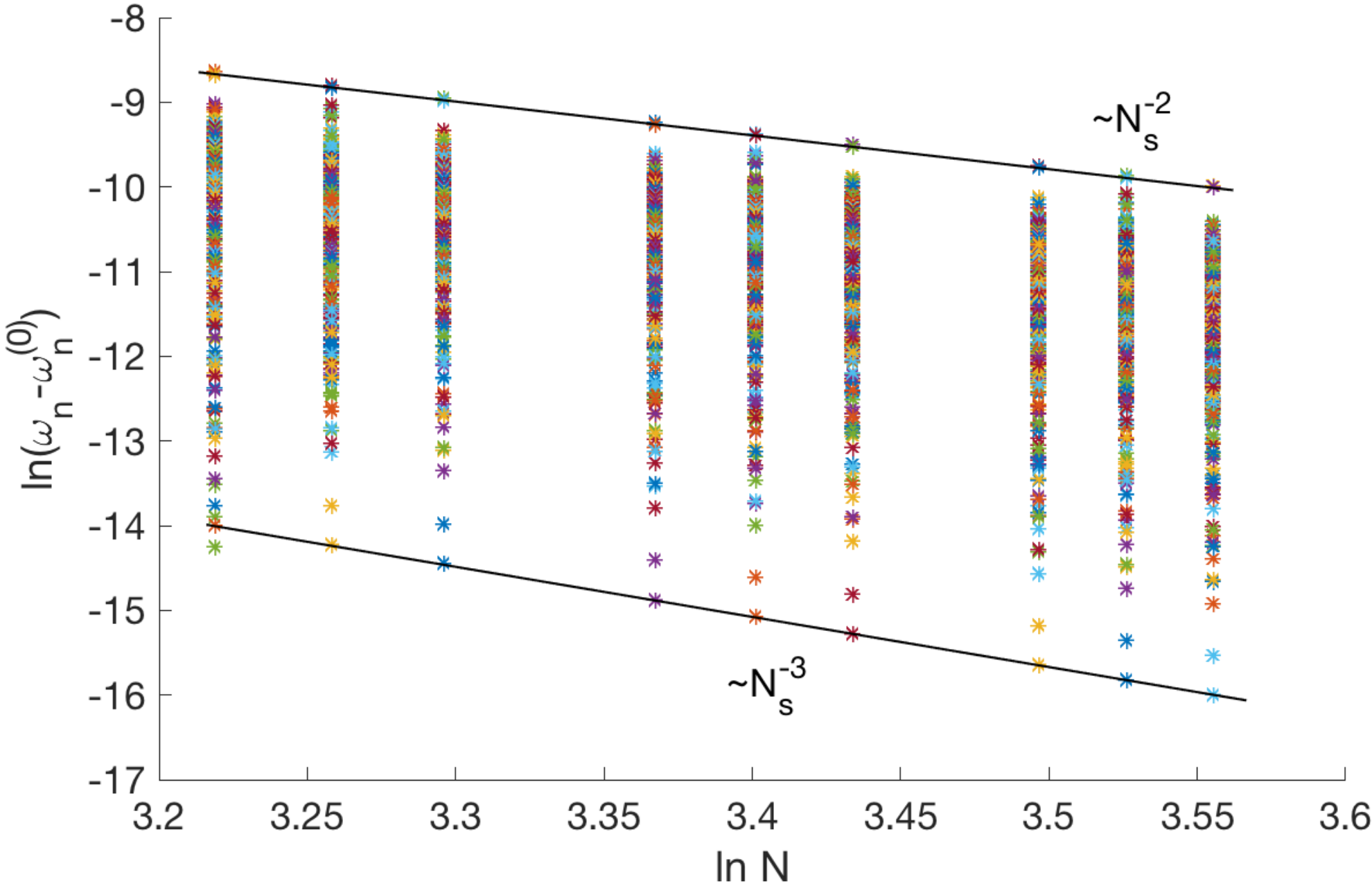}
\caption{Difference between the eigenvalues of the Bogoliubov Hamiltonian with and without the addition of the $B$-matrix (see text) as a function of linear number of sites on a torus in presence of a Skyrmion. There appear two different scaling regimes, for higher and lower energy states.}\label{fig:scaling_w}
\end{figure}

The powerful constraints from analyticity can already be seen at the level of harmonic fluctuations due to $H_2$,
when considering the scaling of $\delta \omega_{n}$ at fixed $\omega_{n}^{(0)}$ as $N$ becomes large. 
The motivation for fixing $\omega_{n}^{(0)}$ is twofold. First, this quantity appears explicitly in the second order
correction to eigenfrequencies, Eq.~(\ref{2_order_correction}). Second, $\omega_{n}^{(0)}$ controls
the ratio between the characteristic scale associated to spatial variations of $| u_{n}^{(0)}\rangle$ and the lattice spacing. Note that, to keep $\omega_{n}^{(0)}$ fixed, we have to choose $n$ proportional to $N$, if eigenvalues are listed in increasing order.
On Fig.~\ref{fig:scaling_w}, we observe that $\delta \omega_{n}$ scales like $N^{-2}$ at large $\omega_{n}^{(0)}$, and like $N^{-3}$ at small $\omega_{n}^{(0)}$. The former scaling is the one suggested by the second order perturbation theory in $B$ for any smooth texture, see Eq.~(\ref{2_order_correction}). The extra decay of  $\delta \omega_{n}$ at large $N$ for small $\omega_{n}^{(0)}$ can be attributed to cancellations in the $B$ term arising from the analytic nature of a minimal energy Skyrmion configuration. More precisely, we can establish that, for a holomorphic texture, 
$\sum_{j \in \mathcal{N}(i)}B_{ij}$ goes to zero faster than $a^{2}$ as $a\rightarrow0$.  

To show this, we project the spin unit sphere onto the complex $w$ plane, using stereographic projection,
so $w=\frac{\sin \theta}{1+ \cos \theta}e^{i\phi}$. From Eq.~(\ref{approx_Bij}), we may write:    
\begin{equation}
\sum_{j\in \mathcal{N}(i)} B_{ij}  \simeq  \frac{JS}{4} \sum_{j\in \mathcal{N}(i)}
\left((\gradbf \theta+i\sin \theta \gradbf \phi)\dotbf(\rbf_{i}-\rbf_{j})\right)^{2}.
\end{equation}
But since $\gradbf \theta+i\sin \theta \gradbf \phi = \frac{2}{1+|w|^{2}}\frac{\bar{w}}{|w|}\gradbf w$, this becomes:
\begin{equation}
\sum_{j\in \mathcal{N}(i)} B_{ij}  \simeq  JS \left(\frac{\bar{w}_{i}}{(1+|w_{i}|^{2})|w_{i}|}\right)^{2}
\sum_{j\in \mathcal{N}(i)} \left( \gradbf w \dotbf (\rbf_{i}-\rbf_{j}) \right)^{2}.
\end{equation}
For a holomorphic texture, $\gradbf w \dotbf (\rbf_{i}-\rbf_{j}) = \frac{\partial w}{\partial z}(\rbf_{i})(z_{i}-z_{j})$.
But for a regular square or triangular lattice:
\begin{equation}
\sum_{j\in \mathcal{N}(i)} (z_{i}-z_{j})^{2}=0,
\end{equation}
which establishes our claim. 

At low energy $\omega_{n}^{(0)}$, the unperturbed eigenstates $| u_{n}^{(0)}\rangle$ of $A$ are slowly varying on the lattice scale, so we can neglect the difference between amplitudes at neighboring sites. As a result,
when we apply the $B$ operator to $| u_{n}^{(0)}\rangle$, the statement in the previous paragraph
implies that the leading $a^{2}$ term vanishes. Inserting this into the second order expression for $\delta \omega_{n}$ provides there an extra  $a^{2}$ prefactor, i.e. it replaces the $N^{-2}$ scaling by  $N^{-3}$. These
two different scalings are visible on Fig.~\ref{fig:scaling_w}.

\section{Generalization to K\"ahler manifolds}
\label{Generalized_to_Kaehler}
In this section, we discuss the question of how to lift the results on vanishing zero-point energies to a more general class of models. This proceeds via the identification of the K\"ahlerian manifold nature underpinning the Skyrmion results and emergence of BPS states. This in particular extends the quantum Hall Skyrmion results to the  $SU(4)\sim CP^3$ Skyrmion excitations of graphene at the charge neutrality point. The following subsections contain material some of which is rather mathematical; the reader mainly interested in further implications of the results in the previous section to spiral states may skip ahead to the next section. 

\subsection{The Hessian for classical sigma models into K\"ahler manifolds}

Here we shall mostly be considering maps from a two-dimensional plane into a K\"ahler manifold $\mathcal{M}$~\cite{Arnold},\cite{GH}, i.e.~$\mathcal{M}$ is a complex manifold
with local complex coordinates $w_i$, equipped with a Hermitian metric
\begin{equation}
ds^{2}=\sum_{ij}h_{ij}dw_{i}d\bar{w}_{j},
\end{equation} 
such that the corresponding associated $(1,1)$ form 
\begin{equation}
\omega=\frac{i}{2}\sum_{ij}h_{ij}dw_{i} \wedge d\bar{w}_{j}
\end{equation}
is closed. This implies that, locally, the metric can be expressed using a K\"ahler potential $\Phi$, specifically
\begin{equation}
h_{ij}=\frac{\partial^{2} \Phi}{\partial w_i \partial \bar{w}_{j}}
\end{equation}
The K\"ahler condition also means that $\mathcal{M}$ is a symplectic manifold, so that it can be viewed as a classical phase-space.

In the geometric quantization procedure, this manifold becomes a parameter space for a family of quantum coherent states which are in some intuitive sense localized in $\mathcal{M}$ in the vicinity their associated parameter~\cite{Perelomov}. Sigma models with K\"ahler manifolds as target spaces have been studied for a long time. For a review, see~\cite{Perelomov87}.

Let us first present a short list of definitions and properties which are useful for our purposes.  
The classical energy functional for a map from the two-dimensional plane into a target space $(x,y)\rightarrow (w_i)$ reads
\begin{equation}
E=\frac{g}{2} \int d^{2}\mathbf{r}\: h_{ij}(w(\mathbf{r}),\bar{w}(\mathbf{r}))\nabla w_i.\nabla\bar{w}_j
\end{equation}
As usual, it is convenient to introduce $z=x+iy$, so that, for any pair of functions $f$ and $g$ the standard
scalar product of their gradients can be written as
\[\nabla f .\nabla g = 2(\partial_{z}f \partial_{\bar{z}}g +  \partial_{\bar{z}}f \partial_{z}g) \]
Therefore, the classical energy functional can also be written in terms of complex coordinates on the plane as
\begin{equation}
E=g \int d^{2}\mathbf{r}\: h_{ij} 
(\partial_{z}w_i \partial_{\bar{z}}\bar{w}_j +  \partial_{\bar{z}}w_i \partial_{z}\bar{w}_j).
\end{equation}
The topological charge density for a classical map $f(z)=(w_i)$ is defined by
\[Q=\int d^{2}\mathbf{r} \: f^{*}\omega. \]
Explicitely,
\begin{equation}
Q=\int d^{2}\mathbf{r}\: h_{ij} 
(\partial_{z}w_i \partial_{\bar{z}}\bar{w}_j -  \partial_{\bar{z}}w_i \partial_{z}\bar{w}_j).
\end{equation}
Due to the fact that the form $\omega$ is closed $d\omega=0$, it is easy to show that $Q$ does not change to a first order under any infinitesimal variation of the map $f$, so $Q$ depends only on the homotopy class of $f$. In many interesting situations, $Q$ takes only integer values. 

An essential property of this class of models is the Bogomolnyi inequality, which reads:
\begin{equation}
E \geq g |Q| 
\label{Bog_inequality}
\end{equation}
To prove it, simply notice that $E=g(A+B)$ and $Q=A-B$, with:
\[A=\int d^{2}\mathbf{r}\: h_{ij} 
\partial_{z}w_i \partial_{\bar{z}}\bar{w}_j,\]
and
\[B=\int d^{2}\mathbf{r}\: h_{ij} \partial_{\bar{z}}w_i \partial_{z}\bar{w}_j.\]
Because the hermitian metric $h_{ij}$ is positive definite, $A$ and $B$ are both positive real numbers, so $A+B \geq |A-B|$, which
gives~(\ref{Bog_inequality}). Suppose that we are interested in configurations with $Q \geq 0$. Then the lower bound~(\ref{Bog_inequality})
is reached for configurations such that $B=0$. Using again the assumption that  $h_{ij}$ is positive definite, this holds only if     
$\partial_{\bar{z}}w_i=0$, i.e. the map $z\rightarrow w_i$ is holomorphic. Likewise, for $Q \leq 0$, minimal configurations are anti-holomorphic.

As a result, there is a massive continuous degeneracy of configurations with a fixed topological charge. Let us assume that $Q \geq 0$ and let us pick such minimal configuration, for which $\partial_{\bar{z}}w_i=0$. We are interested in the behavior of the Taylor expansion of the $E$ functional around this minimum. For this, we replace $w_i$ by $w_i + \delta w_i$. Note that $\delta w_i$ is not assumed
to be holomorphic, contrary to $w_i$. The existence of the Bogomolnyi lower bound induces the following constraint, namely that Taylor expansion of $E$ to arbitrary orders does not contain terms involving only $\delta w_i$ and its derivatives, nor only
$\delta \bar{w}_i$ and its derivatives. In other words, all terms in the Taylor expansion of $E$ involve at least some combinations
of $\delta w_i$ and  $\delta \bar{w}_j$ and their derivatives. To prove this, we use the fact that $B=0$ for the holomorphic reference
configuration. Then, $B$ reads:
\begin{equation}
B=\int d^{2}\mathbf{r}\: h_{ij} \partial_{\bar{z}}\delta w_i \partial_{z} \delta \bar{w}_j
\label{expansion_for_B}
\end{equation}
Here, we see that the Taylor expansion of $B$ begins at second order, and that it contains only mixed terms in $\delta w_i$ and  $\delta \bar{w}_j$ and their derivatives.
This proves the desired statement for $E$ because $E=gQ+2gB$ and the {\em topological} fact that $Q$ does not change under any smooth variation of the configuration.

\subsection{$\mathrm{Gr(M,N)}$ coherent states} 

Let $\mathcal{F}(M,N)$ denote the finite-dimensional Fock subspace of $M$ fermions with $N$ internal states. It is convenient to introduce fermionic creation and annihilation
operators $\hat{f}^{\dagger}_{i}$, $\hat{f}_j$, ($1 \leq i,j \leq N$), which obey standard anticommutation rules $\{f^{\dagger}_{i},f_j\}=\delta_{ij}$. It is well known that the family of Slater
determinants provides an over-complete basis in  $\mathcal{F}(M,N)$, which behaves in many ways as the family of coherent states for a bosonic 
system~\cite{Perelomov,Berezin75,Berezin78,Rowe80,Fujii96}. The Slater determinants for $M$ fermions are defined, up to a global phase, by the $M$-dimensional        
subspace in $\mathbb{C}^{N}$ spanned by the occupied single particle states. The family of such subspaces is known as the 
Grassmannian manifold $\mathrm{Gr(M,N)}$~\cite{GH}. 

Let us pick a reference Slater determinant $\left | \mathcal{S} \right  \rangle = f^{\dagger}_{1}...f^{\dagger}_{M}\left | 0 \right \rangle$. 
To any $(N-M,M)$ matrix $W$ with complex entries $w_{ij}$, we associate the subspace $s(W)$ spanned by the vectors $\mathbf{e}_j + \sum_{i=1}^{N-M}w_{ij}\mathbf{e}_{M+i}$, 
($1\leq j \leq M$), where $(\mathbf{e}_1,...,\mathbf{e}_N)$ is the canonical basis of $\mathbb{C}^{N}$. The subspace $s(W)$ defines 
a Slater determinant $\left | \mathcal{S}(W) \right  \rangle$ by:
\begin{equation}
\left | \mathcal{S}(W) \right  \rangle = \prod_{j=1}^{M}(f^{\dagger}_{j}+\sum_{i=1}^{N-M}w_{ij}f^{\dagger}_{M+i})\left | 0 \right \rangle. 
\end{equation}
An easy calculation shows that:
\begin{equation}
\left \langle \mathcal{S}(W) | \mathcal{S}(W') \right \rangle = \mathrm{Det}(I+W^{\dagger}W').
\label{Slaterdetoverlap} 
\end{equation}

Note that when the $W$ matrix varies, the subspaces $s(W)$ do not cover the whole Grassmannian manifold but a large open subset of it, whose complement is of 
zero measure. It turns out that $\mathrm{Gr}(M,N)$ is a K\"ahler manifold~\cite{GH}. On the open subset where $w_{ij}$ coordinates are defined, the K\"ahler
potential $\Phi$ can be chosen as: 
\begin{equation}
\Phi=\frac{1}{\pi}\log \mathrm{Det} (I+W^{\dagger}W)
\label{defPhi}
\end{equation}
The Berry curvature $\mathcal{B}$ associated to this family of coherent sates is closely related to the symplectic form $\omega$
derived from the K\"ahler potential $\Phi$ on $\mathrm{Gr}(M,N)$. Indeed:
\begin{equation}
\mathcal{B}=2\pi \omega
\label{Berry_curvature}
\end{equation}
To see this, we have to consider first the Berry connection $\mathcal{A}$ defined by:
\begin{equation}
i\mathcal{A}=d_{w} \frac{\left \langle \mathcal{S}(W') | \mathcal{S}(W) \right \rangle}
{(\left \langle \mathcal{S}(W) | \mathcal{S}(W) \right \rangle\left \langle \mathcal{S}(W') | \mathcal{S}(W') \right \rangle)^{1/2}}
\end{equation}
where the differential is taken at $W'=W$.
From the expression for the overlap~(\ref{Slaterdetoverlap}) and the definition~(\ref{defPhi}) of the K\"ahler potential, we get:
\begin{equation}
\mathcal{A}=\frac{\pi}{2i}\left(\frac{\partial \Phi}{\partial w_{ij}}dw_{ij}-\frac{\partial \Phi}{\partial \bar{w}_{ij}}d\bar{w}_{ij}\right)
\end{equation}
The Berry curvature $\mathcal{B}$ is simply the two-form $d\mathcal{A}$, and we immediately get~(\ref{Berry_curvature}). 

A very nice feature of coherent states is that they form an overcomplete basis with a natural representation of the identity operator,
which reads~\cite{Fujii96}: 
\begin{equation}
I= \int_{\mathrm{Gr}(M,N)} d\mu(W) \: \frac{\left | \mathcal{S}(W) \right \rangle\left \langle \mathcal{S}(W) \right |}
{\left \langle \mathcal{S}(W) | \mathcal{S}(W) \right \rangle}
\end{equation}
Here, the measure $d\mu(W)$ is proportional to the volume form $\Omega$ on $\mathrm{Gr}(M,N)$ defined by:
\[\Omega = \frac{\wedge^{M(N-M)}\omega}{(M(N-M))!}\]
Explicitely:
\begin{equation}
\Omega=\left(\mathrm{Det} (I+W^{\dagger}W)\right)^{-N}\prod_{ij}\frac{d\Re w_{ij} \wedge d\Im w_{ij}}{\pi}
\end{equation}
Before giving the coefficient of $\Omega$ in $d\mu(W)$, let us mention that, in many situations, it is useful to
consider a classical limit for a given family of coherent states. In the case of spin coherent states, it is natural 
to view the spin $S$ of the $SU(2)$ representation as the inverse of an effective Planck's constant $\hbar_{\mathrm{eff}}$, 
because the dimension of the representation is $2S+1$, so we may associate to each coherent state a phase-space area on the unit sphere 
equal to $4\pi/(2S+1)$, which should be equal to $2\pi \hbar_{\mathrm{eff}}$ according to Bohr's correspondence principle.
The desired classical limit is then obtained by taking $S\rightarrow \infty$. Note that the spin $S$ representation is obtained by taking
the $2S$ times symmetrized tensor product of the fundamental spin 1/2 representation of $SU(2)$. We will follow the same principle here,
and introduce the $k$ times symmetrized tensor product of the  $\mathcal{F}(M,N)$ Fock subspace. In this new space, denoted by $\mathcal{F}(M,N)^{(k)}$,  
the  $k$ times symmetrized tensor products 
$\left | \mathcal{S}(W),k \right \rangle$ of $\left | \mathcal{S}(W) \right \rangle$ provide a family of coherent states over $\mathrm{Gr}(M,N)$.
The overlaps are now equal to:
\begin{equation}
\left \langle \mathcal{S}(W),k | \mathcal{S}(W'),k \right \rangle = \left(\mathrm{Det}(I+W^{\dagger}W')\right)^{k}.
\end{equation}
The Berry connection $\mathcal{A}$ and curvature $\mathcal{B}$ are multiplied by $k$, and the representation of the identity becomes:
\begin{equation}
I= \int_{\mathrm{Gr}(M,N)} d\mu(W,k) \: \frac{\left | \mathcal{S}(W),k \right \rangle\left \langle \mathcal{S}(W),k \right |}
{\left \langle \mathcal{S}(W),k | \mathcal{S}(W),k \right \rangle}
\end{equation} 
The integration measure is given by~\cite{Fujii96}: 
\begin{equation} 
d\mu(W,k)=\frac{\mathcal{N}(M,k)}{\mathcal{N}(M,N-M+k)}\Omega
\end{equation}
where:
\begin{equation}
\mathcal{N}(M,k)=\frac{0!1!...(M-1)!}{k!(k+1)!...(k+M-1)!}
\end{equation}

As for spins, where the spin $S$ representation of $SU(2)$ is conveniently described in terms of Schwinger bosons, we introduce
a collection of bosonic operators $b_{\alpha}$, $b^{\dagger}_{\alpha}$, where the label $\alpha$ denotes strictly increasing sequences
$1 \leq i_1 < i_2 < ... < i_M \leq N$. With these notations, single boson states $b^{\dagger}_{\alpha}|0>$ form a basis of  
$\mathcal{F}(M,N)$. Generalizing, it is clear that $\mathcal{F}(M,N)^{(k)}$ is the subspace in the bosonic Fock space defined by the
constraint:
\begin{equation}   
\sum_{\alpha} b^{\dagger}_{\alpha}b_{\alpha}=k
\label{k_constraint}
\end{equation}
Let us now consider the expectation value of normal-ordered products of bosonic creation and annihilation operators. We have:
\begin{widetext}
\begin{equation}
\frac{\left \langle \mathcal{S}(W),k |\prod_{\alpha} (b^{\dagger}_{\alpha})^{m_{\alpha}}(b_{\alpha})^{n_{\alpha}}|\mathcal{S}(W),k \right \rangle}
{\left \langle \mathcal{S}(W),k | \mathcal{S}(W),k \right \rangle}=\frac{k!}{(k-n)!}
\frac{\prod_{\alpha}\mathcal{M}_{\alpha}(\bar{W})^{m_{\alpha}}\mathcal{M}_{\alpha}(W)^{n_{\alpha}}}
{\left(\mathrm{Det}(I+W^{\dagger}W)\right)^{n}}
\label{basic_cov_symbol}
\end{equation}
\end{widetext}
In this formula, we have assumed that $\sum_{\alpha}m_{\alpha}=\sum_{\alpha}n_{\alpha}=n$.
The functions $\mathcal{M}_{\alpha}(W)$ are simply the coefficients of $\left| \mathcal{S}(W) \right \rangle$ on the Fock basis, i.e.:
\begin{equation}
\left| \mathcal{S}(W) \right \rangle=\sum_{\alpha}\mathcal{M}_{\alpha}(W)b^{\dagger}_{\alpha}\left|0\right\rangle
\end{equation}
The $\mathcal{M}_{\alpha}(W)$ function is explicitely given by the minor determinant associated to rows $i_1,i_2,...,i_M$ of the $N$ by $M$ matrix
$V=\left(\begin{array}{c}I_{M} \\ W \end{array} \right)$, so it is a polynomial in $w_{ij}$'s.   

We are now in a position to generalize a result obtained earlier~\cite{DKM16} in the case of Schwinger boson representations of $SU(N)$ spins,
corresponding to geometric quantization above the complex projective space $\mathbb{C}P(N-1)$. Suppose that we have an operator $\hat{H}$ acting on
$\mathcal{F}(M,N)^{(k)}$, such that its expectation value on coherent states $\left | \mathcal{S}(W),k \right \rangle$ has a minimum at $W=0$.
If the Taylor expansion of this expectation value around $W=0$ contains only mixed terms in $w_{ij}$'s and $\bar{w}_{ij}$'s (or equivalently, no term
involving only $w_{ij}$'s or only $\bar{w}_{ij}$'s), then $\left | \mathcal{S}(0),k \right \rangle$ is the exact ground-state of 
$\hat{H}$. The proof is based on eq.~(\ref{basic_cov_symbol}) and is very similar to the one given earlier~\cite{DKM16} 
for Schwinger boson representations of $SU(N)$ spins.

\subsection{Generalization to K\"ahler manifolds}

Here, $\mathcal{M}$ is a K\"ahler manifold, for which a geometric quantization exists in the sense given by Berezin~\cite{Berezin74}.
In this case, the quantum Hilbert space $\mathcal{H}$ is realized as the space of holomorphic sections of a positive complex line bundle
$L$ over $\mathcal{M}$. Coherent states can be defined as follows. Suppose that $w_i$ are complex coordinates on a dense open subset of 
 $\mathcal{M}$ whose complement has zero measure. Then to any quantum state $\left | \Psi \right\rangle$ we associate the holomorphic
function $\psi(w)$. Let us consider an orthonormal basis $\left | \chi_{\alpha} \right\rangle$ in $\mathcal{H}$, with corresponding
holomorphic functions $\chi_{\alpha}(w)$. The completeness relation reads:
\be
\psi(w)=\sum_{\alpha}\chi_{\alpha}(w) \left\langle\chi_{\alpha}|\Psi\right\rangle 
\ee
Introducing the state $\left| \mathcal{C}(\bar{w})\right\rangle$ defined by:
\be
\left| \mathcal{C}(\bar{w}) \right\rangle = \sum_{\alpha}\chi_{\alpha}(w)^{\ast} \left|\chi_{\alpha}\right\rangle 
\ee
we see that, for any state $\left | \Psi \right\rangle$:
\be
\psi(w)=\left\langle \mathcal{C}(\bar{w})|\Psi\right\rangle 
\ee
This implies that if $\psi(w)=0$, then $\left| \mathcal{C}(\bar{w})\right\rangle$ is orthogonal to $\left | \Psi \right\rangle$. But the condition
$\psi(w)=0$ defines an hyperplane in $\mathcal{H}$ whose orthogonal line is generated by $\left| \mathcal{C}(\bar{w})\right\rangle$. Intuitively, this
can be interpreted by saying that $\left| \mathcal{C}(\bar{w})\right\rangle$ is a state localized near $(w_i)$ in phase-space $\mathcal{M}$.   

As for the case $\mathcal{M}=\mathrm{Gr}(M,N)$ already considered, it is convenient to introduce alternative quantizations by taking the $k$ times symmetrized tensor product of
$\mathcal{H}$, denoted by $\mathcal{H}^{(k)}$. As before, the limit $k \rightarrow \infty$ is
interpreted as a classical limit. It is then straightforward to introduce a Schwinger boson
representation of $\mathcal{H}^{(k)}$, where to each basis state  
$\left|\chi_{\alpha}\right\rangle$ we associate a pair of canonically conjugate bosonic operators
$b^{\dagger}_{\alpha}$, $b_{\alpha}$. Then $\mathcal{H}^{(k)}$ is here again the subspace in the bosonic Fock space defined by the constraint:
\begin{equation}   
\sum_{\alpha} b^{\dagger}_{\alpha}b_{\alpha}=k
\end{equation}
 
In  $\mathcal{H}^{(k)}$, generalized coherent states are defined by:
\be
\left| \mathcal{C}(\bar{w}),k \right\rangle = \left(\sum_{\alpha}\chi_{\alpha}(w)^{\ast} 
b^{\dagger}_{\alpha}\right)^{k}\left| 0 \right\rangle
\ee

The basic equation~(\ref{basic_cov_symbol}) can now be generalized as:
\begin{widetext}
\begin{equation}
\frac{\left \langle \mathcal{C}(\bar{w}),k |\prod_{\alpha} (b^{\dagger}_{\alpha})^{m_{\alpha}}(b_{\alpha})^{n_{\alpha}}|\mathcal{C}(\bar{w}),k \right \rangle}
{\left \langle \mathcal{C}(\bar{w}),k | \mathcal{C}(\bar{w}),k \right \rangle}=\frac{k!}{(k-n)!}
\frac{\prod_{\alpha}\chi_{\alpha}(w)^{m_{\alpha}}(\chi_{\alpha}(w)^{\ast})^{n_{\alpha}}}
{\left\langle \mathcal{C}(\bar{w})|\mathcal{C}(\bar{w})\right\rangle^{n}}
\label{gen_basic_cov_symbol}
\end{equation}
\end{widetext}

We can then prove as before that, for any operator $\hat{H}$ acting on
$\mathcal{H}^{(k)}$, such that its expectation value on coherent states $\left | \mathcal{C}(\bar{w}),k \right \rangle$ has a minimum at $w=0$, and if
the Taylor expansion of this expectation value around $w=0$ contains only mixed terms in $w_{i}$'s and $\bar{w}_{j}$'s (or equivalently, no term
involving only $w_{i}$'s or only $\bar{w}_{j}$'s), then $\left | \mathcal{C}(0),k \right \rangle$ is the exact ground-state of $\hat{H}$.

\section{Zero-point fluctuations in spirals}

This section addresses the situation regarding zero-point fluctuations in a spiral magnet. 
We consider the limit in which the pitch of the spiral is fixed while taking the lattice spacing to zero. 
In doing so, we scale the coupling constants of the lattice model such that the total energy of the spiral over one period is fixed.  As in the above
examples, we find zero-point fluctuations to vanish in the continuum limit. 

In the first subsection, we present an explicit calculation 
for the well-studied  one-dimensional case of a Dzyaloshinksii-Moriya spiral. The second subsection presents a family of  Hamiltonians for which the vanishing of the zero-point energy is exact already at the lattice scale, 
and to which the Dzyaloshinksii-Moriya spiral is proximate in the sense of joining that family in the continuum limit.

\subsection{Dzyaloshinksii-Moriya spiral}
The one-dimensional spiral we consider arises for a spin-orbit coupled Hamiltonian
\begin{equation}\label{DM_ham}
{\cal H}^{\mathrm{DM}}=\sum_i J {\bf S}_i\cdot{\bf S}_{i+1}+D \hat{z}\cdot({\bf S}_i\times{\bf S}_{i+1})\ ,
\end{equation}
where $D$ denotes the strength of the Dzyaloshinksii-Moriya interaction. The classical ground state for this interaction has the spins residing in the $xy$-plane perpendicular to the DM-vector $\hat{z}$. Successive spins are twisted at an angle $\theta$ given by $\tan\theta=D/J$, so that the wavelength of the spiral is inversely proportional to $D$. 

\begin{figure}[t]
\centering
\includegraphics[width=0.95\columnwidth]{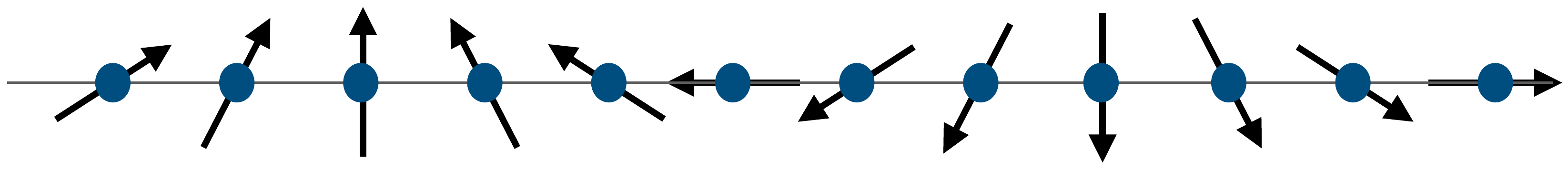}
\caption{Schematic picture of a spin-spiral state with spins rotating by angle $\theta$ on successive lattice sites.}\label{fig:spin-spiral}
\end{figure}

First we perform a rotation of the Hamiltonian into a basis which is locally co-aligned with the classical spin configuration,
see the general discussion in previous sections. In the present case the spin transformation simplifies to
\begin{figure}[b]
\centering
\includegraphics[width=0.95\columnwidth]{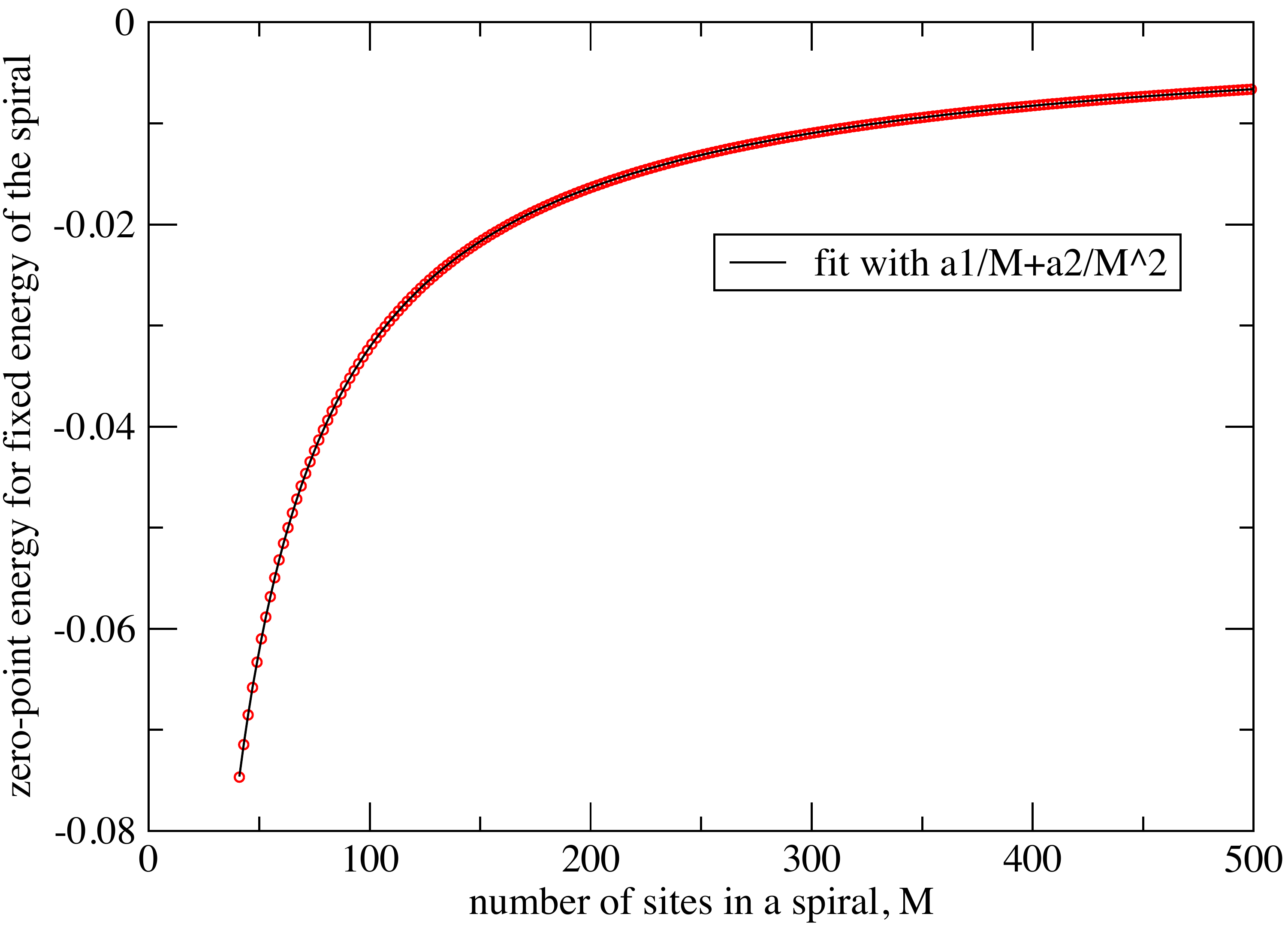}
\caption{Scaling of zero-point energy for a DM-spiral as a function of a number of lattice sites $M$ in a single period. Note that we perform our calculations in the thermodynamic limit with $n\times M$ lattice sites where integer $n\gg 1$. (red dots) Zero point energy obtained from numerical diagonalization of the Hamiltonian $\hat{\mathcal{H}}_2^{\mathrm{DM}}$ in the background of a classical spiral spin-texture. (solid line) Fit with $c_1/M+c_2/M^2$. Note that the coupling strength is scaled to keep the energy of the spiral fixed.}\label{fig:scaling_DM_spiral}
\end{figure}

\begin{equation*}
\hat{S}^{+}_{i}=[\hat{T}^z_i+\frac{1}{2}(\hat{T}^{+}_i-\hat{T}^{-}_{i})]e^{i\phi_i},\ 
\hat{S}^{z}_{i}=-\frac{1}{2}(\hat{T}^{+}_i+T^{-}_{i}),
\end{equation*}
where $\phi_i$ is the angle between axis $x$, and the spin on site $i$, and we fix the angle between two neighbouring spins such that $\theta=\phi_{i+1}-\phi_{i}=2\pi/M$, where $M$ is the number of sites in the period of a spiral. 
The value of $D/J$ is then obtained as $D/J=\arctan(2\pi/M)$.

After substituting this transformation into the Hamiltonian (\ref{DM_ham}), and performing Holstein-Primakoff transformation to  first order in $S$ we obtain
\begin{multline}\label{eq:HP_DM}
\hat{\mathcal{H}}^{\mathrm{DM}}_2=E_0+\frac{\varepsilon_0}{2}\sum_{m}[\hat{b}^{\dagger}_m \hat{b}_m-\cos^2(\theta/2)\hat{b}^{\dagger}_m \hat{b}_{m+1}\\+\sin^2(\theta/2)\hat{b}_m\hat{b}_{m+1}+h.c.],
\end{multline}
where $\varepsilon_0=|J|/\cos\theta$, and $E_0=-N \varepsilon_0/4$ for the system with $N$ lattice sites. In the continuum limit the classical energy of the spiral is $\sim |J|M\theta^2/2$, and we have to scale $J$ as $const\times M$ to make this energy independent of $M$ for large $M$. In this limit the coefficient in front of the Bogoliubov term scales as $\sin^2(\theta/2)\sim1/M^2$ and the zero-point energy (which has another prefactor of $\varepsilon_0$) vanishes in the continuum limit as $1/M$, see Fig. 6.

It is instructive to rewrite the bosonic Hamiltonian in momentum space. We consider a lattice with $N$ sites and periodic boundary conditions. We introduce bosonic operators in momentum space, and Fourier transform the Hamiltonian using $\hat{b}_m=1/\sqrt{N}\sum_{m=1}^N \hat{b}_k e^{i k m}$, which gives
\begin{multline}
\hat{\mathcal{H}}^{\mathrm{DM}}_2=E_0+\hat{h}_0+\varepsilon_0\sum_{k\ne0}[1-\cos^2(\theta/2)\cos{k}]\hat{b}^{\dagger}_k\hat{b}_k\\+\frac{1}{2}\sin^2(\theta/2)(e^{-ik}\hat{b}_k \hat{b}_{-k} +e^{i k}\hat{b}^{\dagger}_k\hat{b}^{\dagger}_{-k}),
\end{multline}
where  $\hat{h}_0$ is the zero-mode $(k=0)$ Hamiltonian 
\begin{equation}
\hat{h}_0=\varepsilon_0\sin^2(\theta/2)[\hat{b}^{\dagger}_0\hat{b}_0+\frac{1}{2}(\hat{b}_0 \hat{b}_{0} +\hat{b}^{\dagger}_0\hat{b}^{\dagger}_{0})].
\end{equation}
This Hamiltonian can be re-written in an instructive form after introducing a generalized momentum operator $\hat{P}=\hat{b}_0+\hat{b}^{\dagger}_0$, and an effective mass $m^{-1}=\varepsilon_0\sin^2(\theta/2)$.
\begin{equation}
\hat{h}_0=-\frac{\varepsilon_0}{2}\sin^2(\theta/2)+\frac{\hat{P}^2}{2 m},
\end{equation}
where the first term gives the contribution to the zero-point energy of the zero-mode, and the second term corresponds to the kinetic energy related to translations of the system as a whole. Note that the first term scales as $1/M$ in the continuum limit.

After introducing Bogoliubov transformations
\begin{equation}
\hat{b}_{k}=u_k\hat{\beta}_k+v_k\hat{\beta}^{\dagger}_{-k},
\end{equation}
we  obtain the contribution to zero-point energy from finite momentum modes. Defining
\begin{equation}
\gamma_k=1-\cos^2(\theta/2)\cos k,\ \eta_k=\sin^2(\theta/2)\cos k,
\end{equation}
and $\tanh 2\varphi_k=-\eta_k/\gamma_k$, then
\begin{equation}
u_k=\cosh\varphi_k,\ v_k=\sinh\varphi_k,
\end{equation}
the Hamiltonian can be written in a diagonal form
\begin{equation}
\hat{\mathcal{H}}^{\mathrm{DM}}_2=E_0+\delta+\frac{\hat{P}^2}{2 m}+\sum_{k>0}\omega_k\hat{\beta}^\dagger_k\hat{\beta}_k,
\end{equation}
where the excitation spectrum $\omega_k$ is given by
\begin{equation}
\omega_k=\gamma_k(u^2_k+v^2_k)+2\eta_k u_k v_k,
\end{equation}
and the zero-point energy $\delta$ reads
\begin{equation}
\delta=-\frac{\varepsilon_0}{2}\sin^2(\theta/2)+\varepsilon_0\sum_{k\ne 0} (\gamma_k v^2_k +\eta_k u_k v_k).
\end{equation}

This is shown in Fig.~\ref{fig:scaling_DM_spiral}. We indeed find a nonzero zero-point energy, which  vanishes in the continuum limit
$M\rightarrow\infty$ like a power law, $1/M$. 

\subsection{Generalised ferromagnets} 
We supplement the above considerations by presenting a family of models in which zero-point fluctuations are absent
as they are unitarily related to ferromagnets. To see this, consider  the ferromagnetic lattice Hamiltonian, $\hat{H}$ (Eq.~\ref{spin_hamiltonian}),
and its undressed ground state. We now introduce a unitary transformation, denoted by ${\cal{U}}$, which 
transforms vectors in, and operators on, Hilbert space, respectively as
\begin{eqnarray}
|v\rangle &\rightarrow& {\cal{U}}\, |v\rangle \nonumber \\
O &\rightarrow& {\cal{U}}\, O\, {\cal{U}}^\dagger \label{eq:unit_ferro}
\end{eqnarray}
For the choice of $O=\hat{H}$ and $|v\rangle=|\uparrow\rangle$, the uniformly ferromagnetic groundstate, the resulting
state ${\cal{U}}\, |\uparrow\rangle$ trivially is the undressed ground state of ${\cal{U}}\, \hat{H}\, {\cal{U}}^\dagger$.

As an aside, we note that such a transformation remarkably is not of purely academic interest but has plays an important role in the  study of Kitaev-Heisenberg models, where it reveals an unexpected exactly-soluble point for a complex spin-orbit coupled Hamiltonian.\cite{Jackeli-Khalliulin}

Here, we find use for this transformation as an explanation of the behaviour of the zero-point energy of the DM spiral as we move towards the continuum limit. The basic point is that the DM Hamiltonian can {\it almost} be cast in the form ${\cal{U}}\, \hat{H}\, {\cal{U}}^\dagger$, with ${\cal U}$ a rotation matrix relating neighbouring spins by the pitch angle $\theta$. However, inspecting  Eq.~\ref{eq:HP_DM} shows that its eigenvalues do not have precisely unit modulus. For this to be the case, the normal (hopping) terms would themselves need to be multiplied by a factor which -- crucially -- approaches an identity in the continuum limit $\theta\rightarrow0$.
This would correspond to reducing the first term in ${\cal H}^\mathrm{DM}$ as a function of $D$. It is the failure of Eq.~\ref{eq:HP_DM} to include this correction which leads to the appearance of the zero-point fluctuations for finite $\theta$. 

For clarity, we note that this argument does not apply directly to the Skyrmion textures mentioned above -- there, we use the unrotated ferromagnetic Hamiltonian on a topologically stabilised state representing a local energy minimum.

\section{Conclusion}

We showed that the degeneracy of Bogomolny-Prasad-Sommerfield manifold is not lifted by quantum fluctuations in the case of general non-linear sigma-models with the target space given by K\"ahler manifolds presenting a physically-important example of the case of Grassmanian manifold relevant to quantum Hall effect in graphene. In addition we exposed our theory on a more pedestrian level by considering continuum limits of lattice models, which helps to show that analytic spin textures are  special compared to more general textures in terms of finite-size scaling of zero-point corrections in taking the continuum limit. Further, we showed that the phenomenon of vanishing zero-point motion can appear more generally in slowly-twisted almost ferromagnets, with the DM spiral as a concrete example. 

From a broader perspective, beyond the implications to exotic magnets, we feel this work provides two interesting angles on long-standing interesting issues in statistical physics and field theory more generally. 

One is the existence of undressed states generally. While ferromagnets are known 
to be special, it was perhaps less obvious that twisted magnets also can remain undressed in the continuum limit; and also, that this phenomenology exhibits additional substructure as evidenced by the role of BPS-states in accelerating the vanishing of zero-point fluctuations in the continuum limit, which role can be encountered in the entire class of K\"{a}hler manifolds. 

The other item is the behaviour of `non-universal' quantities, i.e.\ those involving information from the lattice scale, in the `universal' continuum limit. Here, the Casimir energy of zero-point fluctuations vanishes in the continuum limit, but is nonzero for any lattice discretisation. 
This Casimir energy, however, does play a physical role, e.g.\ in the lifting of ground-state 
degeneracies in a process known as quantum order by disorder. 
From this vantage point, our results imply that the energy scale on which quantum order by disorder becomes visible will decrease with the size of a Skyrmion, and will vanish particularly quickly for BPS-Skyrmions. 

Both of these items may perhaps serve as a useful reminder that much of physics, in condensed matter at least, does involve the presence of a finite scale parameter, and that the loss of information which occurs in the continuum limit may not be unimportant in all settings of physical interest. Also, the concrete results obtained here
in the absence of supersymmetry about the absence of zero-point fluctuations may either be a feature entirely unrelated to the more familiar instances arising in relativistic field theory from the cancellation of fluctuations in bosonic and fermionic sectors, or they may be more pedestrian and perhaps intuitively accessible instances of the same physics.  

\section*{Acknowledgements}
This work was in part supported by the Deutsche Forschungsgemeinschaft under grant SFB 1143. D.K.~was supported by EPSRC Grant No.~EP/M007928/2.

\end{document}